\documentclass[prl,twocolumn]{revtex4}
\usepackage{graphicx}

\begin{document}

\author{K. Uhl\'\i\v{r}ov\'a}
\email{klara@mag.mff.cuni.cz}
\author{J. Prokle\v{s}ka}
\author{V. Sechovsk\'y}
\affiliation{Charles University, DCMP, Ke Karlovu 5, Praha 2, 121 16, Czech Republic}

\title{Comment on "Emergence of a Superconducting State from an Antiferromagnetic Phase in Single Crystals of the Heavy Fermion Compound
Ce$_2$PdIn$_8$"}

\maketitle

A recently published Letter~\cite{Kaczorowski2009} has reported on
the antiferromagnetism (AF) and ambient-pressure superconductivity
(SC) in a Ce$_2$PdIn$_8$ single crystal with $T_{\rm N}\sim 10$~K
and $T_{\rm c} = 0.68$~K, respectively. Although we very much
appreciate the effort exerted to prepare and characterize this new
heavy fermion (HF) superconductor (SC), we would like to add a
cautionary note that the reported N\'eel temperature coincides
remarkably with $T_{\rm N} =10.2$~ K of
CeIn$_3$~\cite{Lawrence1980}. It therefore leads us to considers
the possible presence of CeIn$_3$ in the samples that were
investigated. In other Ce$_nT$In$_{3n+2} (n = 1,2)$
compounds~\cite{Chen2002,Kim2004,Movshovic2001} the AF is either
absent ($T$ = Co, Ir) or remarkably limited to much lower
temperatures ($T$ = Rh). These compounds form a
quasi-two-dimensional tetragonal structure with the CeIn$_3$ and
$T$In$_2$ layers alternating along the $(001)$ direction. Hence
one might expect that the AF correlations develop within the
CeIn$_3$ layers while the interaction between the layers will be
weaker as reported for CeRhIn$_5$, an incommensurate AF ($T_{\rm
N} = 3.8$~K)~\cite{Bao2000}. The remarkable agreement of the
$T_{\rm N}$ values in the reported Ce$_2$PdIn$_8$ with the
well-known CeIn$_3$ is not discussed in the
Letter~\cite{Kaczorowski2009}. Neither the striking discrepancy
between their own results on single
crystals~\cite{Kaczorowski2009} and polycrystals (reported
paramagnetic down to 0.35K~\cite{KaczorowskiPB}) has been
explained. The absence of SC in the polycrystalline sample is
explained by an unconventional coupling sensitive to structural
disorder, internal strains, and/or tiny changes in the
composition, but the disagreement in the magnetic ground state is
not discussed at all.

Although a detailed phase analysis (X-ray diffraction and
micro-probe) of the crystals was claimed to have been
done~\cite{Kaczorowski2009}, we would, however, still like to
suggest that a CeIn$_3$ single crystal covered by a
single-crystalline layer of Ce$_2$PdIn$_8$ was in fact that which
was investigated. From the reported heat capacity data, we
estimate the amount of CeIn$_3$ to be $15-20\%$. In such case, a
microprobe analysis of the sample's surface would not be able to
detect it. Also, most of the diffraction peaks of both compounds
interfere, because they have an almost equal lattice parameter $a
= 0.4693$~nm~~\cite{Shtepa08} and $a = 0.4689$~nm~\cite{Harris1965} for
Ce$_2$PdIn$_8$ and CeIn$_3$, respectively. 

\begin{figure}[b]
\centering
\includegraphics[width=\columnwidth]{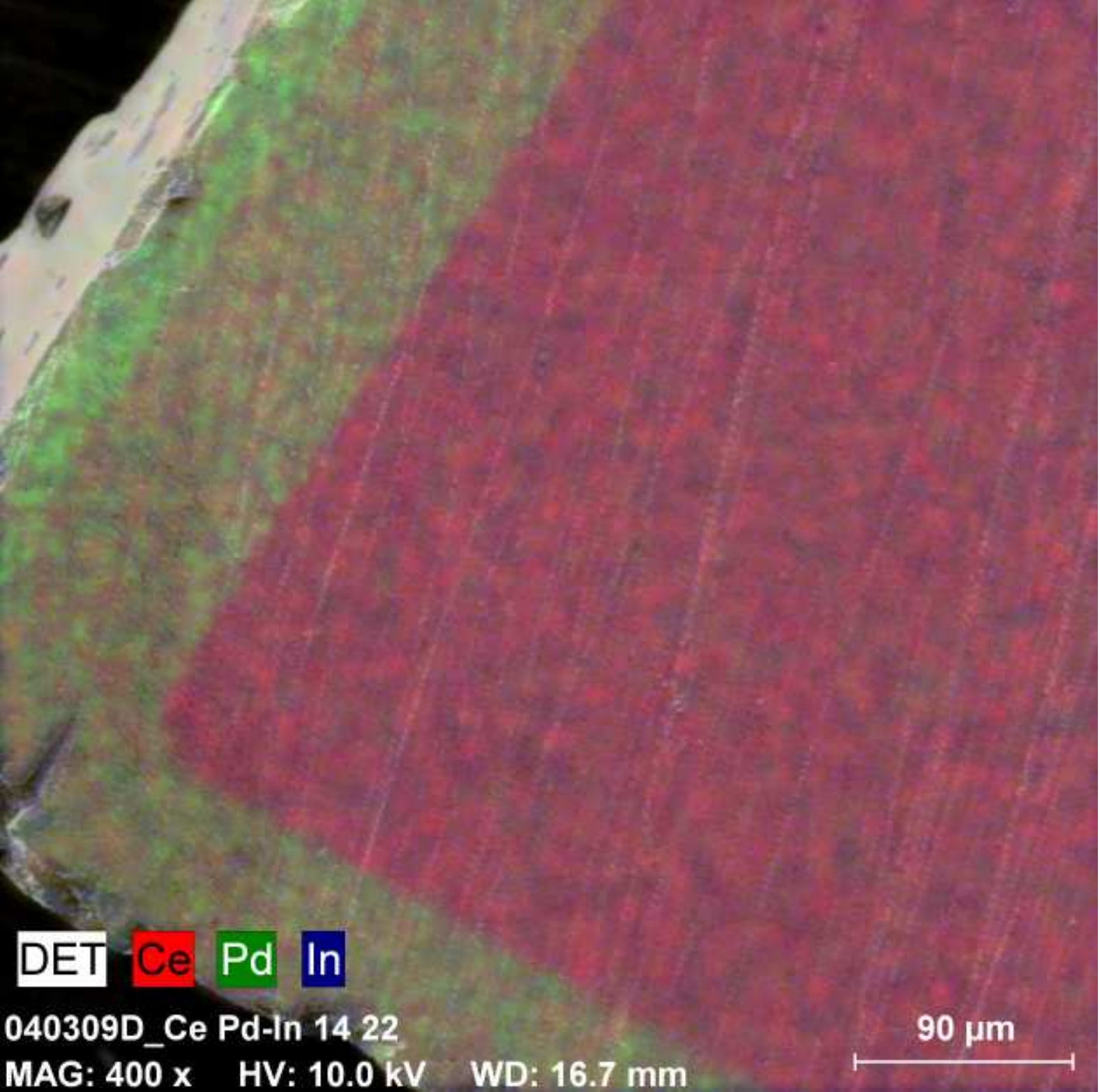}
\caption{
EDX element mapping (Ce-red, Pd-green, In-blue) of a typical polished sample. CeIn$_3$ (red region) is covered by a layer of Ce$_2$PdIn$_8$ and
Ce$_{1.5}$Pd$_{1.5}$In$_7$ (thin dark green).
}\label{fig1}
\end{figure}

Our first magnetization data obtained on crystals grown
analogously to~\cite{Kaczorowski2009} were in agreement with the
Letter. A careful microprobe analysis, however, indicated a
presence of CeIn$_3$, and element mapping showed that
Ce$_2$PdIn$_8$ and CeIn$_3$ form a sandwich-like system with
well-defined regions (see Fig.\ref{fig1}). 

To confirm that the AF originates in CeIn$_3$, we have measured
more than 5 different CeIn$_3$--free samples of Ce$_2$PdIn$_8$ by
means of resistivity, heat capacity and magnetic susceptibility,
and paramagnetic behaviour with significant magnetocrystalline anisotropy was observed down to a SC temperature~\cite{Uhlirova10}.

The SC with $T_{\rm c} =
0.7$~--~$0.45$~K (sample dependent) has been confirmed in our
samples. The difference of critical temperature is probably given
by structural planar defects, which were also observed in
Ce$_2$RhIn$_8$~\cite{Moshopoulou06}. In agreement with
~\cite{Kaczorowski2009}, the SC has a HF character and it is a
bulk property of the compound but it
does not emerge
out of a long-range AF state below the N\'eel temperature of 10 K
because the reported AF was due to presence of an impurity phase.

\end{document}